\documentclass[twocolumn,twoside,slac_two]{revtex4}
\usepackage{graphicx}
\usepackage{fancyhdr}
\begin{document}

\title{An External Inverse Compton Emission Model of Gamma-Ray Burst High-Energy Lags}

%

\author{K. Toma$^{1,2}$, X. F. Wu$^{1,2,3}$, and P. M\'{e}sz\'{a}ros$^{1,2,4}$}
\affiliation{$^1$Department of Astronomy and Astrophysics, Pennsylvania State University, 525 Davey
Lab, University Park, PA 16802, USA}
\affiliation{$^2$Center for Particle Astrophysics, Pennsylvania State University, University Park,
PA 16802, USA}
\affiliation{$^3$Purple Mountain Observatory, Chinese Academy of Sciences, Nanjing 210008, China}
\affiliation{$^4$Department of Physics, Pennsylvania State University, University Park, PA 16802, USA}
%

\begin{abstract}
The {\it Fermi} satellite has been reporting the detailed temporal properties of gamma-ray bursts (GRBs) 
in an extremely broad spectral range, 8 keV - 300 GeV, in particular, the unexpected delays of the 
GeV emission onsets behind the MeV emission of some GRBs. We focus on GRB 080916C, one of the 
{\it Fermi}-LAT GRBs for which the data of the delayed high-energy emission are quite extensive, 
and we show that the behavior of the high-energy emission of this burst can be explained by a model 
in which the prompt emission consists of two components: 
one is the MeV component due to the synchrotron-self-Compton radiation of electrons 
accelerated in the internal shock of the jet and the other is the high-energy component due to 
inverse Compton scattering of the photospheric X-ray emission of the expanding cocoon off the same electrons 
in the jet. Such an external inverse Compton effect could be important for other {\it Fermi}-LAT GRBs, 
including short GRBs as well. In this model, the delay timescale is directly linked to the physical
properties of GRB progenitor. 
\end{abstract}

\maketitle

\thispagestyle{fancy}


\section{Introduction}

Gamma-ray bursts (GRBs) were only sparsely observed in the $> 100$~MeV range, 
until the {\it Fermi} satellite was launched on June 11 2008 \citep{atwood09}.
Now {\it Fermi} provides extremely broad energy coverage, $8~{\rm keV}-300~{\rm GeV}$,
with high sensitivity for GRBs, and is accumulating a wealth of new data which
open a completely new window on the physics of GRBs.
The high-energy temporal and spectral data provided by {\it Fermi} can severely
constrain the physical parameters of the GRB emission region and the circumburst
environment, which will lead to a deeper understanding of the central engine and 
the GRB progenitors, and will also constrain models of high-energy cosmic ray 
acceleration \cite{meszaros06,falcone08}.

GRB 080916C has the largest isotropic $\gamma$-ray energy release so far, 
$E_{\gamma,{\rm iso}} \simeq 8.8 \times 10^{54}\;$erg (with redshift $z\simeq 4.35$).
{\it Fermi} LAT obtained its high-energy emission data quite extensively, showing
several important new properties \cite{abdo09}:
\begin{itemize}
\item[(i)]
The time-resolved spectra (with resolution $\sim 5-50$~s) are well fitted by a 
smoothly broken power-law function (the so-called Band function)
from 8~keV up to a photon with energy $\approx 13.2$~GeV.
\item[(ii)]
The $\varepsilon > 100$~MeV emission is not detected together with the first 
$\varepsilon \lesssim 1$~MeV pulse and the onset of the $\varepsilon>100$~MeV 
emission coincides with the rise of the second pulse ($\approx 5$~s after the trigger).
\item[(iii)]
Most of the emission in the second pulse shifts towards later times as higher 
energies are considered.
\item[(iv)]
The $\varepsilon>100$~MeV emission lasts at least 1400~s, while photons with 
$\varepsilon<100$~MeV are not detected past 200~s.
\end{itemize}
Some other {\it Fermi}-LAT GRBs also display
high-energy lags, similar to the properties (ii) and/or (iii) \cite{abdo09,abdo09b}, 
and then they should be very important to understand the prompt emission mechanism of GRBs.
We will call the $\varepsilon \lesssim 1$~MeV emission and 
the $\varepsilon>100$~MeV emission "MeV emission" and "high-energy emission", 
respectively.

A simple physical picture for the property (i) is that the prompt emission consists of 
a single emission component, such as synchrotron radiation of electrons accelerated
in internal shocks of a relativistic jet.
In this picture, the peak of the MeV pulse could be attributed to the cessation of
the emission production (i.e., the shock crossing of the shell) and the 
subsequent emission could come from the high latitude regions of the shell.
Thus the observed high-energy lag for the second pulse (property (iii)) requires
that the electron energy spectrum should be harder systematically in the higher
latitude region.
This would imply that the particle acceleration process should definitely depend on 
the global parameters of the jet, e.g., the angle-dependent relative Lorentz factor 
of the colliding shells, but such a theory has not been formulated yet.
The property (ii) could be just due to the fact that the two pulses originate in
two internal shocks with different physical conditions for which the electron
energy spectrum of the second internal shock is harder than that of the first one
\cite{abdo09}.

Another picture is that the prompt emission consists of the MeV component and 
a delayed high-energy component.
The latter component could be produced by hadronic effects (i.e., photo-pion
process and proton synchrotron emission) \cite{asano09,razzaque09}, 
but they require extremely large total energy budget \cite{wang09}.

\begin{figure}
\includegraphics[width=65mm]{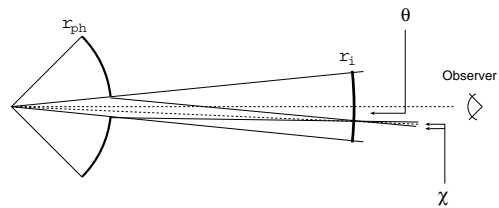}
\caption{
Geometry of our model.
The jet with opening angle $\theta_j \simeq 0.01$ and the cocoon with 
opening angle $\theta_c \simeq 0.8$ are ejected from the collapsing star.
The cocoon becomes optically thin at $r = r_{\rm ph}$ some time ($\lesssim 5$~s) after the burst trigger.
The second pulse of the prompt emission is produced by the accelerated electrons
in the internal shock of the jet at $r=r_i$, and the cocoon X-ray photons
are up-scattered by the same electrons at $r=r_i$ into the GeV range.
}
\label{fig:geometry}
\end{figure}

In this paper, we discuss a different two-component emission picture in which
the delayed high-energy component is produced by leptonic process 
(i.e., electron inverse Compton scattering).
We focus on the effect that the ambient radiation up-scattered by the accelerated
electrons in the jet can have a later peak than that of the synchrotron and 
synchrotron-self-Compton (SSC) emission of the same electrons
(corresponding to the property (iii))
\cite{wang06,fan08,beloborodov05}.
Provided that the seed photons for the Compton scattering come from the region
behind the electron acceleration region of the jet (see Figure~\ref{fig:geometry}), 
the up-scattered high-energy photon field is highly anisotropic in the comoving
frame of the jet, i.e., the emissivity is much larger for the head-on collisions of 
the electrons and the seed photons. 
As a result, a stronger emission is observed from the higher latitude regions,
and thus its flux peak lags behind the synchrotron and SSC emission.

Here we propose that the seed photons may be provided by the photospheric 
emission of an expanding cocoon.
GRB 080916C is a long GRB, and it may originate from the collapse of a massive 
star.
The relativistic jet produced in the central region penetrates the star and deposits
most of its energy output into a thermal bubble, or cocoon,
until it breaks out of the star \cite{meszaros01,matzner03,wzhang03,morsony07}.
The cocoon can store an energy comparable or larger than the energy of
the prompt emission of the jet, and thus it may make an observable signature 
outside the star \cite{ramirez02,peer06}.
The cocoon escaping from the star will emit soft X-rays, and these can be 
up-scattered by the accelerated electrons in the jet into the high-energy range.
The optical thinning of the expanding cocoon may be delayed behind the prompt
emission of the jet, so that the onset of the high-energy emission is delayed
behind the MeV emission (corresponding to the property (ii)).
Thus this model has the potential for explaining the two delay timescales;
the delayed onset of the high-energy photons (property (ii))
is due to the delayed emission of the cocoon,
while the high-energy lag within the second pulse (property (iii)) is due to
the anisotropic inverse Compton scattering.
We also show that the combination of the time-averaged spectra of the SSC and
the up-scattered cocoon (UC) emission is roughly consistent with the observed
smooth power-law spectrum (property (i)) (see Figure~\ref{fig:spectrum}).

As we will explain below, the UC emission is short-lived and may not
account for the whole high-energy emission, which lasts much longer than
the MeV emission (property (iv)).
It is natural that the high-energy emission in later times is related to
the afterglow, i.e., produced by the external shock which propagates in the 
circumburst medium. This possibility is studied by \cite{kumar09}.
They claim that even the onset phase of the high-energy emission is produced
by the external shock. However, the rise of the flux of GRB 080916C in the LAT
energy range ($\sim t^6$) is too steep for the external shock to reproduce it.
Thus at least the first part of the delayed high-energy component 
of this burst should be related to the prompt emission.

Although we focus on GRB 080916C to examine whether the UC emission is viable 
for its properties in the high energy range in this paper, 
our modeling is generic, and the UC emission
could be important for some other {\it Fermi}-LAT GRBs (including short GRBs).
We will discuss the temporal and spectral properties of the cocoon emission
(in \S~2), the synchrotron and SSC emission of the jet (in \S~3),
and the UP emission of the jet (in \S~4). We summarize our model and
its implications in \S~5. For more details, see \cite{toma09}.

\section{Cocoon Emission}
\label{sec:cocoon_emission}

We take the total energy and the total mass of the cocoon and the stellar radius
as parameters, $E_c$, $M_c$, and $r_*$, respectively.
After the jet breaks out of the star, the cocoon expands with the sound speed,
$c_s = c/\sqrt{3}$, where $c$ is the speed of light.
The cocoon expands by its own thermal pressure in the comoving frame 
as expected in the standard fireball model \cite{meszaros06}.
Its expansion speed in the comoving frame suddenly becomes close to $c$, and then
the opening angle of the cocoon measured in the central engine frame is given by
$\theta_c \simeq \Gamma_{c,{\rm ex}}^{-1}$, where $\Gamma_{c,{\rm ex}}$ is the 
Lorentz factor corresponding to $c_s$.
Thus we obtain $\theta_c \simeq \sqrt{2/3} \simeq 0.8$.
This rough estimate is consistent with the results of the numerical simulations \cite{morsony07}.

The cocoon material accelerates as $\Gamma \propto r$ and 
reaches the terminal Lorentz factor $\Gamma_c = E_c/M_c c^2$ at
$r_s = r_* \Gamma_c = 5.0\times10^{12}~r_{*,11} (\Gamma_c/50)~$cm.
(We adopt the notation $Q = 10^x Q_x$ in cgs units throughout this paper.)
For $r>r_s$, the cocoon material is dominated by the kinetic energy.
The radiation stored in the cocoon is released when the opacity for the electron
scattering becomes less than unity.
The photosphere radius is given by
\begin{eqnarray}
r_{\rm ph} &\simeq& 
\left[\frac{E_c \sigma_T}{2\pi (1-\cos\theta_c) \Gamma_c m_p c^2}\right]^{1/2} \nonumber \\
&\simeq& 2.1 \times 10^{14}~E_{c,52}^{1/2} 
\left(\frac{\Gamma_c}{50}\right)^{-1/2}~{\rm cm},
\label{eq:photospheric_radius}
\end{eqnarray}
where $\sigma_T$ is Thomson cross section.

The cocoon may become optically thin later than the onset of the first MeV pulse
(i.e., the burst trigger time).
Let $t=0$ be the photon arrival time at the earth emitted at the stellar surface
at the jet breakout.
The first MeV pulse produced within the jet at $r=r_i$ is observed at
$t \simeq \Delta t_i \equiv r_i(1+z)/(2c \Gamma_j^2)$,
where $\Gamma_j$ is the bulk Lorentz factor of the jet and $z$ is the source 
redshift.
This timescale is comparable to the angular spreading timescale of the pulse,
and we can take $\Delta t_i \simeq 2$~s for GRB 080916C.
The second MeV pulse is observed $\simeq 5$~s after the burst trigger, i.e.,
at $t \simeq \Delta t_i + 5 \simeq 7$~s.
On the other hand, the cocoon photospheric emission is observed at
\begin{equation}
t \simeq \Delta t_c \equiv \frac{r_{\rm ph}}{2c \Gamma_c^2} (1+z)
\simeq 7.5~E_{c,52}^{1/2} \left(\frac{\Gamma_c}{50}\right)^{-5/2}~{\rm s},
\label{eq:delta_t_c}
\end{equation}
where we have used $z \simeq 4.35$.
The cocoon photospheric emission may be observed from $t=\Delta t_c$ to 
$t = \Delta t_c + \Delta t_d$, where $\Delta t_d \sim \Delta t_c$ is the time during
which the cocoon will be adiabatically cooled.
If internal dissipation occurs in the jet at $r = r_i > r_{\rm ph}$ making the 
second MeV pulse within the duration of the cocoon emission, 
the cocoon photons may be up-scattered to higher energies by the energetic 
electrons within the dissipation region of the jet, 
which may be observed along with the second MeV pulse.
Therefore we require a condition $\Delta t_i < \Delta t_c \lesssim \Delta t_i + 5$.
This condition puts a constraint to the physical parameters of the cocoon,
\begin{equation}
0.3 < E_{c,52}^{1/2} \left(\frac{\Gamma_c}{50}\right)^{-5/2} \lesssim 0.9
\label{eq:co_constraint1}
\end{equation}
We adopt the parameters 
$E_c \simeq 10^{52}$~erg and $\Gamma_c \simeq 50$
for the purposes of calculating the flux of the cocoon emission.

The comoving temperature of the cocoon when its opening angle becomes $\theta_c$ is approximately
given by $T'_{\rm init} \simeq [E_c/(2\pi (1-\cos\theta_c) r_*^3 a)]^{1/4}$, where $a$ is the 
Stefan constant. Then the temperature at the photosphere is estimated to be
$T'_{\rm ph} = T'_{\rm init} (r_s/r_*)^{-1} (r_{\rm ph}/r_s)^{-2/3}$.
Non-thermal electrons injected by internal shocks within the photosphere may
make the emission be quasi-thermal. Its spectum is written by
\begin{equation}
F^{\rm co}_{\varepsilon} = F^{\rm co}_{\varepsilon_{\rm ph}} \times \left\{
\begin{array}{lc}
\left(\frac{\varepsilon}{\varepsilon^{\rm co}_{\rm ph}}\right)^2  
& {\rm for}~\varepsilon < \varepsilon^{\rm co}_{\rm ph}, \\
\left(\frac{\varepsilon}{\varepsilon^{\rm co}_{\rm ph}}\right)^{\beta} 
& {\rm for}~ \varepsilon^{\rm co}_{\rm ph} < \varepsilon < 
\varepsilon^{\rm co}_{\rm cut},
\end{array}
\right.
\label{eq:cocoon_spectrum}
\end{equation}
where $\varepsilon_{\rm ph}^{\rm co}$ and $F_{\varepsilon_{\rm ph}}^{\rm co}$ are 
given by
\begin{equation}
\varepsilon_{\rm ph}^{\rm co} \simeq 2.82~ kT'_{\rm ph} \frac{2\Gamma_c}{1+z}
\simeq 1.2~E_{c,52}^{-1/12} r_{*,11}^{-1/12}~{\rm keV} 
\left(\frac{\Gamma_c}{50}\right)
\label{eq:cocoon_nu}
\end{equation}
\begin{eqnarray}
F_{\varepsilon_{\rm ph}}^{\rm co} & \simeq & \frac{(1+z)^3}{d_L^2} 
\frac{2\pi (\nu_{\rm ph}^{\rm co})^2}{c^2} kT'_{\rm ph} 
\Gamma_c \left(\frac{r_{\rm ph}}{\Gamma_c}\right)^2 \nonumber \\
& \simeq & 31~E_{c,52}^{3/4} r_{*,11}^{-1/4}~{\rm keV}~{\rm cm}^{-2}~{\rm s}^{-1}~{\rm keV}^{-1},
\label{eq:cocoon_fnu}
\end{eqnarray}
where $\nu_{\rm ph}^{\rm co} = \varepsilon_{\rm ph}^{\rm co}/h$ ($h$ is the Planck
constant), and we have taken the luminosity distance of GRB 080916C as 
$d_L \simeq 1.2 \times 10^{29}$~cm.
Some numerical calculations of the radiative processes in the cocoon show
$\beta \sim -1$ and $\varepsilon_{\rm cut}^{\rm co} \sim 30 \times \varepsilon_{\rm ph}^{\rm co}$
\cite{peer06}.

The observation of GRB 080916C shows that there is no excess from the Band spectrum
at the X-ray band, $\gtrsim 10$~keV, and we obtain a rough upper limit of the
cocoon X-ray emission 
$\varepsilon_{\rm ph}^{\rm co} F_{\varepsilon_{\rm ph}}^{\rm co} \lesssim
40~{\rm keV}~{\rm cm}^{-2}~{\rm s}^{-1}$.
This limit leads to another constraint on the cocoon parameters,
\begin{equation}
r_{*,11} \gtrsim 0.8~E_{c,52}^2 \left(\frac{\Gamma_c}{50}\right)^3.
\label{eq:co_constraint3}
\end{equation}

\section{Synchrotron and SSC Emission}
\label{sec:emission_formalism}

We can constrain, from the {\it Fermi} observation, the global physical parameters of the 
jet: the bulk Lorentz factor $\Gamma_j$, the emission radius of the second pulse $r_i$, 
and the opening angle $\theta_j$.
First of all, from the absence of a $\gamma\gamma$ absorption cutoff,
we obtain a lower limit on the bulk Lorentz factor of the jet, 
$\Gamma_j \gtrsim 870$ \citep{abdo09}.
Since the angular spreading timescale of the pulse is $\Delta t_i \simeq 2$~s
for the second pulse, similar to the first pulse, the emission radius is
estimated by
\begin{equation}
r_i \simeq 2c \Gamma_j^2 \frac{\Delta t_i}{1+z} 
\simeq 2.2 \times 10^{16}~\Gamma_{j,3}^2 
\left(\frac{\Delta t}{2~{\rm s}}\right)~{\rm cm}.
\end{equation}

Since GRB 080916C is so bright, it is probable that the jet is viewed on-axis,
and we adopt this assumption as a simplification (see Figure~\ref{fig:geometry}).
In this case, the cocoon is viewed off-axis, since the jet cone will not be
filled with the cocoon material.
The cocoon emission is thus less beamed, but this off-axis effect is not
significant because the opening angle of the jet can be estimated to be small.
The isotropic $\gamma$-ray energy of this burst is $8.8 \times 10^{54}$~erg.
To obtain a realistic value of the collimation-corrected $\gamma$-ray energy,
$\lesssim 10^{51}$~erg, the jet opening angle is constrained by 
$\theta_j \lesssim 0.015$.
We adopt $\theta_j \simeq 0.01$, and having adopted a nominal value of 
$\Gamma_c \simeq 50$ in accord with the observed time delay of the high-energy emission 
(equation \ref{eq:co_constraint1}), we obtain
\begin{equation}
\Gamma_c \theta_j \simeq 0.5 < 1.
\end{equation}
Thus the off-axis dimming and softening effects are not significant for
the cocoon emission.

We assume that the jet is dominated by the kinetic energy of protons 
and we estimate the physical parameters 
of the jet dissipation region for the second MeV pulse.
In the scenario where the dissipation is due to internal shocks \cite{meszaros06},
the collisionless shock waves can amplify the magnetic field and 
accelerate electrons to a power-law energy distribution, which then 
produce synchrotron radiation and SSC radiation.
At $r=r_i$, the comoving number density of the jet is estimated by
$n'= L \Delta t_i/(4\pi r_i^3 m_p c^2 (1+z))$, where $L$ is the isotropic-equivalent luminosity of the jet.
The internal energy density produced by the internal shock is given by $u' = n' \theta_p m_p c^2$,
where $\theta_p$ is a factor of order unity.
Assuming that a fraction $\epsilon_B$ of the internal energy of the protons 
is carried into the magnetic field, the field strength is estimated by 
$B' = (8\pi \epsilon_B u')^{1/2}$.
Assuming that a fraction $\epsilon_e$ of the internal energy of the protons is
given to the electrons, the minimum Lorentz factor of the electrons is given by
$\gamma_m = [(p-2)/(p-1)](m_p/m_e)\epsilon_e \theta_p$, where $p$ is the index of
the electron energy distribution and we have assumed that it is similar to the value
inferred from the spectrum of the first pulse, $p \simeq 3.2$.
We find that the cooling electron Lorentz factor $\gamma_c \sim \gamma_m$ for 
fitting the observational data. 
In this case we obtain $\gamma_c \simeq 360 (\epsilon_B/10^{-5})^{-1/3} (\tau/4\times10^{-4})^{-2/3}$,
where $\tau = \sigma_T n' r_i/\Gamma_j$ is the Thomson optical depth.

The synchrotron characteristic energy and the synchrotron peak flux 
(at the synchrotron energy corresponding to $\gamma_c$) are estimated by
\begin{eqnarray}
\varepsilon_m &\simeq& \frac{3h e B'}{4\pi m_e c}\gamma_m^2 \frac{2\Gamma_j}{1+z} 
\simeq 2.7~L_{55}^{1/2} \Gamma_{j,3}^{-2} \left(\frac{\Delta t_i}{2~{\rm s}}\right)^{-1} \nonumber \\
&\times& \theta_p^{5/2} 
\left(\frac{\epsilon_B}{10^{-5}}\right)^{1/2} \left(\frac{\epsilon_e}{0.4}\right)^2~{\rm eV},
\label{eq:syn_num}
\end{eqnarray}
\begin{eqnarray}
F_{\varepsilon_c} &\simeq& \frac{\sqrt{3} e^3 B' N}{m_e c^2} 
\frac{2\Gamma_j (1+z)}{4\pi d_L^2} 
\simeq 1.3 \times 10^4~L_{55}^{3/2} \Gamma_{j,3}^{-3} \nonumber \\
&\times& \theta_p^{1/2} \left(\frac{\epsilon_B}{10^{-5}}\right)^{1/2}
~{\rm keV}~{\rm cm}^{-2}~{\rm s}^{-1}~{\rm keV}^{-1}~
\label{eq:syn_fnuc}
\end{eqnarray}
where $N = [L \Delta t_i/(1+z)]/(\Gamma_j m_p c^2)$.
The 1st-order SSC characteristic energy and the SSC peak flux are approximately
\begin{eqnarray}
\varepsilon_m^{\rm SC} &\simeq& 4\gamma_m^2 \varepsilon_m \simeq
1.7~L_{55}^{1/2} \Gamma_{j,3}^{-2} \left(\frac{\Delta t_i}{2~{\rm s}}\right)^{-1} \nonumber \\
&\times& \theta_p^{9/2} 
\left(\frac{\epsilon_B}{10^{-5}}\right)^{1/2} \left(\frac{\epsilon_e}{0.4}\right)^4~{\rm MeV},
\label{eq:SSC_num}
\end{eqnarray}
\begin{eqnarray}
F_{\varepsilon_c}^{\rm SC} &\simeq& \tau F_{\varepsilon_c} 
\simeq 3.4~L_{55}^{5/2} \Gamma_{j,3}^{-8} \left(\frac{\Delta t_i}{2~{\rm s}}\right)^{-1} \nonumber \\
&\times& \theta_p^{1/2} \left(\frac{\epsilon_B}{10^{-5}}\right)^{1/2}
~{\rm keV}~{\rm cm}^{-2}~{\rm s}^{-1}~{\rm keV}^{-1}.
\label{eq:SSC_fnuc}
\end{eqnarray}
We find that the 1st-order SSC radiation can account for the observed MeV emission of this burst.
The emission just below $\varepsilon_c$ is suppressed by synchrotron self-absorption effect,
and the 2nd-order SSC emission is suppressed by the Klein-Nishina effect.

\section{Upscattered Cocoon (UC) emission}
\label{sec:aic}

Here we derive the observed spectrum of the UC emission as a function of the 
polar angle $\theta$ of the emitting region on the shell 
(see Figure~\ref{fig:geometry}). 
The spectrum of radiation scattered at an angle $\theta'_{\rm sc}$ relative to
the direction of the photon beam in the Thomson scattering regime is given by
\citep{brunetti01,wang06,fan08}
\begin{equation}
j'_{\varepsilon'}(\theta'_{\rm sc}) = \frac{3}{2}
\sigma_T (1-\cos\theta'_{\rm sc})\int d\gamma N'(\gamma) \int^1_0 dy
J'_{\varepsilon'_s}(1-2y+2y^2),
\label{eq:aic_emissivity}
\end{equation}
where $y=\varepsilon'/[2\gamma^2 \varepsilon'_s (1-\cos\theta'_{\rm sc})]$.
This is the scattered radiation emissivity in the jet comoving frame, 
$N'(\gamma)$ is the electron energy spectrum, 
and $J'_{\varepsilon'_s}$ is the intensity of the seed photons averaged over solid angle,
i.e., the mean intensity.
The term $\xi \equiv 1-\cos\theta'_{\rm sc}$ describes the anisotropy of the
spectrum, and this is due to the fact that the IC scattering is strongest
for the head-on collisions between electrons and seed photons.
This implies that the UC emission in the observer frame is stronger from the 
high-latitude region of the shell, so that its flux peak lags behind the
onset of the synchrotron and SSC emission of the same electrons, which have
isotropic energy distribution in the comoving frame of the jet.

In order to concentrate on the time-averaged spectrum including the high-latitude
emission, we calculate the flux of the UC emission 
by neglecting the radial structure of the emitting shell for simplicity.
The peak energy and the peak flux of the UC emission are reduced as functions
of the angle parameter $q(\theta) \equiv \Gamma_j^2\theta^2$:
\begin{eqnarray}
\varepsilon_m^{\rm UC} &=& 2\gamma_m^2 \varepsilon_{\rm ph}^{\rm co} 
\frac{\xi(\theta)}{1+\Gamma_j^2 \theta^2} \nonumber \\
&\simeq& 160~{\rm MeV}~\left[\frac{4 q}{(1+q)^2}\right]
\left(\frac{\gamma_m}{400}\right)^2 
\left(\frac{\varepsilon_{\rm ph}^{\rm co}}{1~{\rm keV}}\right), 
\label{eq:co_nu} \\
\varepsilon_m^{\rm UC} F_{\varepsilon_m}^{\rm UC} &=& 3 \tau \gamma_m \gamma_c
\varepsilon_{\rm ph}^{\rm co} F_{\varepsilon_{\rm ph}}^{\rm co}  
\frac{\xi^2(\theta)}{[1+\Gamma_j^2\theta^2]^3}
\nonumber \\
&\simeq& 580~{\rm keV}~{\rm cm}^{-2}~{\rm s}^{-1}~
\left[\frac{40 q^2}{(1+q)^5}\right] \left(\frac{\tau}{4\times10^{-4}}\right) \nonumber \\
&\times&  \left(\frac{\gamma_m}{400}\right)
\left(\frac{\gamma_c}{400}\right)
\left(\frac{\varepsilon_{\rm ph}^{\rm co} F_{\varepsilon_{\rm ph}}^{\rm co}}
{30~{\rm keV}{\rm cm}^{-2}{\rm s}^{-1}}\right),
\label{eq:co_fnu}
\end{eqnarray}
where the functions in the brackets $[~]$ both have values of zero at $q=0$ and 
$q=\infty$ and have peaks of $1$ at $q=1$ and $\simeq 1.4$ at $q=2/3$, respectively.
This means that the UC flux has a peak at $q \simeq 1$, or 
$\theta \simeq \Gamma_j^{-1}$, i.e.,
the peak time of the UC emission lags behind that of the SSC emission on the 
angular spreading timescale, $\Delta t_i \simeq 2$~s.
This is consistent with the observed lag of the GeV emission onset behind 
the MeV emission peak of the second pulse of GRB 080916C.
Here the values of the jet parameters $\tau = 4\times10^{-4}$ and $\gamma_m
=\gamma_c=400$ are applicable for the 1st-order SSC emission of the jet being 
consistent with the observed MeV emission component (see \S~2).
This indicates that the UC emission of the electrons accelerated in the internal 
shock of the jet, emitting the observed MeV emission, can naturally explain 
the observed flux in the GeV range.

\begin{figure}
\includegraphics[width=65mm]{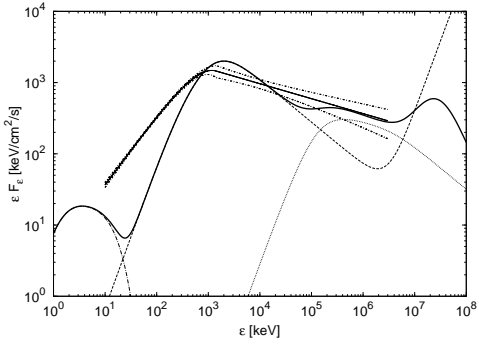}
\caption{
Time-averaged spectrum of the second pulse calculated 
in the up-scattered cocoon emission model.
The 1st-order SSC component plus 2nd-order SSC component without including the
Klein-Nishina effect ({\it dashed line}), the cocoon photospheric emission
({\it dot-dashed line}), and the UC emission ({\it dotted
line}) are shown.
The {\it thick solid line} represents the combination of these, taking account of 
the Klein-Nishina effect, which is roughly consistent, at $\varepsilon \gtrsim 1$~MeV,
with the Band model spectrum ({\it thin solid line}) with 95\% confidence errors 
({\it dot-short-dashed lines}) (from the LAT/GBM group of {\it Fermi}).
The bump at $\sim 30$~GeV is so dim as not to be detected.
The adopted parameters are listed in equation (\ref{eq:cocoon_parameter}) and 
(\ref{eq:jet_parameter}).
}
\label{fig:spectrum}
\end{figure}

If the flux of the cocoon photospheric X-rays is given, i.e., $E_c$, $\Gamma_c$, and
$r_*,$ are given, the fluxes of the UC and SSC emission of the jet
are determined by the jet parameters $L, \Gamma_j, \Delta t_i, 
\epsilon_B$, and $\epsilon_e$.
Since $\Delta t_i \sim 2$~s is roughly given by the 
observations, and this value is necessary
to explain the observed high-energy lag timescale, we have four free parameters.
On the other hand, we have four characteristic observables; the peak fluxes and 
peak photon energies of the SSC component and the UC component.
Therefore the jet parameters are expected to be constrained tightly.

\begin{figure}
\includegraphics[width=65mm]{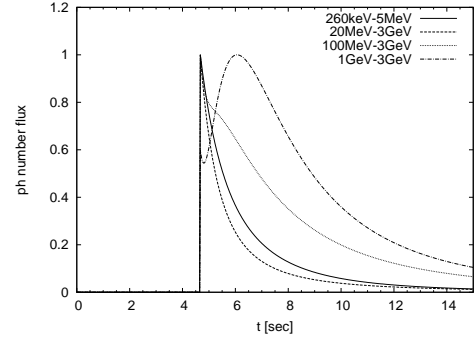}
\caption{
Photon number fluxes of the second pulse in several frequency ranges calculated in the up-scattered
cocoon emission model.
Each lightcurve is normalized to a peak flux of unity.
The peak of the GeV lightcurve is delayed behind that of the MeV lightcurve.
}
\label{fig:lightcurve1}
\end{figure}

Figure~\ref{fig:spectrum} shows the result of the time-averaged spectrum of
the second pulse for the cocoon parameters
\begin{equation}
E_{c,52} = 1.0,~~ \Gamma_c = 52,~~ r_{*,11} = 2.5,
\label{eq:cocoon_parameter}
\end{equation}
and $\beta = -1.2$.
These values satisfy the constraints on the cocoon parameters, equations
(\ref{eq:co_constraint1}) and (\ref{eq:co_constraint3}).
This figure shows that our model is roughly consistent with the observed spectrum
at $\varepsilon \gtrsim 1$~MeV.
The adopted values of the jet parameters are
\begin{equation}
L_{55} = 1.1,~~ \Gamma_{j,3} = 0.93,~~ \Delta t_i = 2.3~{\rm s},~~
\epsilon_B = 10^{-5},~~\epsilon_e = 0.4,
\label{eq:jet_parameter}
\end{equation}
and $p = 3.2$.
The corresponding values of the optical depth for electron scattering and 
the characteristic electron Lorentz factors are
$\tau = 3.5 \times 10^{-4}, \gamma_m = 400, \gamma_c = 390$.
Figure~\ref{fig:lightcurve1} shows the results of the multi-band lightcurves
for the same parameters.
Each lightcurve is normalized to a peak flux of unity.
This clearly displays the lag of the high-energy emission peak.

\section{Discussion}
\label{sec:summary}

We have discussed a model in which the prompt emission spectrum consists of an
MeV component produced by the SSC emission of electrons accelerated in 
internal shocks in the jet and the high-energy component produced by up-scattering 
of the cocoon X-rays off the same electrons (UC emission), and we have shown that this model can explain 
the above three observed properties (i), (ii), and (iii) listed in \S~1.
The expanding cocoon may become optically thin some time later than the first internal
shock of the jet (equation \ref{eq:delta_t_c}), so that the first MeV pulse may not
be associated with the UC emission while the second MeV pulse
may be associated with it (property (ii)).
The UC emission has an anisotropic energy distribution in the comoving frame of the
jet so that the observed UC emission is stronger from the higher-latitude region
of the shell.
This results in the lag of the flux peak of the UC emission behind the MeV emission
onset on the angular spreading timescale (property(iii)).
Figure~\ref{fig:spectrum} shows that the combination of the SSC and UC emission
can reproduce the observed high-energy spectral data 
(property (i)).
The UC emission is short-lived (roughly for $\Delta t_c \simeq 7.5\;$s) 
and may not account for the whole high-energy
emission which lasts longer than the MeV emission (property (iv)).
It is natural that the high-energy emission in the later times is related to the 
afterglow. This has been shown by \cite{kumar09}. However, the early portion of
the high-energy emission should be the UC emission, because the external shock
cannot reproduce the observed steep rise of the flux.

We have focused on GRB 080916C because of its extensive data on the high-energy emission.
However our model is generic, and it could apply
to other {\it Fermi}-LAT GRBs with typical parameters 
$L \simeq 10^{53}~{\rm erg}~{\rm s}^{-1}$, $\theta_j \simeq 0.1$,
and $\Gamma_j \simeq 300$ for which the prompt emission in the soft
$\gamma$-ray range is produced by the 1st-order SSC radiation of
electrons \cite{toma09}.
Some short GRBs might originate from the collapses of the massive stars
\citep{zhang09}, and thus they could have the delayed UC emission.
Even if other short GRBs are produced by the compact star mergers,
it might be possible that the jet is accompanied by the delayed disc wind
\citep{metzger08} and that the emission from the disc wind is 
up-scattered by the electrons accelerated in the jet.
For either progenitor models, the delayed high-energy emission 
associated with short GRBs, if any, would provide an interesting
tool to approach their origins.

A simple prediction of our model is that prompt emission spectra of 
some GRBs would have an excess above the Band spectrum around $\sim 1$~keV 
due to the cocoon photospheric emission, and this excess should have a different 
temporal behavior from that of the MeV emission.
In addition, we expect GRB 080916C to have had bright synchrotron emission 
in the optical band, like the "naked-eye" GRB 080319B.

If our model is correct, we can constrain the parameter range for which
hadronic effects are important on the high-energy emission of GRBs, and 
we can also constrain the models of high-energy cosmic ray acceleration.
Also, the delay time of the onset of the high-energy emission
is directly linked to the optical-thinning time of the expanding cocoon, which
constrains the physical parameters of the progenitor star and the cocoon material
of GRBs.
For GRB 080916C, the stellar radius $r_*$ and the total energy $E_c$ and mass $M_c$
of the cocoon are constrained to be 
$0.3 < E_{c.52}^{-2} (M_c/10^{-4}M_{\odot})^{2.5} \lesssim 0.9$ and 
$r_{*,11} \gtrsim 0.8~E_{c,52}^5 (M_c/10^{-4}M_{\odot})^{-3}$
(see equations \ref{eq:co_constraint1} and \ref{eq:co_constraint3}). 
The cocoon energy and the cocoon mass come from the jet energy 
released within the star and the stellar mass swept by the jet, respectively.
These constraints therefore provide potential tools for investigating the 
structure of the progenitor star just before the explosion, as well as the physical 
conditions of the jet propagating inside the stellar envelope through either 
analytical \cite{meszaros01,matzner03} or numerical \cite{wzhang03,morsony07} approaches.

\begin{acknowledgments}
We acknowledge NASA NNX09AT72G, NASA NNX08AL40G, and NSF PHY-0757155 for partial support.
XFW was supported by the National Natural Science Foundation of China
(grants 10503012, 10621303, and 10633040), National Basic Research Program
of China (973 Program 2009CB824800), and the Special Foundation for the 
Authors of National Excellent Doctorial Dissertations of P. R. China by
Chinese Academy of Sicences.
\end{acknowledgments}









%

\bigskip 

\end{document}